\documentclass[twocolumn,floatfix,amsmath,amssymb,groupedaddress,noeprint]{revtex4-1}

\usepackage{graphicx}
\usepackage{braket}
\usepackage{color}
\usepackage{wrapfig}
\usepackage[margin=1 in]{geometry}

\newcommand{\sixJ}[6] {\left\{\begin{array}{ccc} #1 & #2 & #3 \\ #4 & #5 & #6 \end{array}\right \}}
\newcommand{\threeJ}[6]{\left(\begin{array}{ccc} #1 & #3 & #5 \\ #2 & #4 & #6 \end{array}\right )}

\begin{document}

\title{A Scalable Quantum Computing Platform Using Symmetric-Top Molecules}

\author{Phelan Yu}
\email{phelan@cua.harvard.edu}
\affiliation{Harvard-MIT Center for Ultracold Atoms, Cambridge, MA 02138, USA}
\affiliation{Department of Physics, Harvard University, Cambridge, MA 02138, USA}
\author{Lawrence W. Cheuk}
\affiliation{Harvard-MIT Center for Ultracold Atoms, Cambridge, MA 02138, USA}
\affiliation{Department of Physics, Harvard University, Cambridge, MA 02138, USA}
\author{Ivan Kozyryev}
\altaffiliation{Current address: Department of Physics, Columbia University, New York, NY 10027}
\affiliation{Harvard-MIT Center for Ultracold Atoms, Cambridge, MA 02138, USA}
\affiliation{Department of Physics, Harvard University, Cambridge, MA 02138, USA}
\author{John M. Doyle}
\affiliation{Harvard-MIT Center for Ultracold Atoms, Cambridge, MA 02138, USA}
\affiliation{Department of Physics, Harvard University, Cambridge, MA 02138, USA}

\begin{abstract}
We propose a new scalable platform for quantum computing (QC) -- an array of optically trapped symmetric-top molecules (STMs) of the alkaline earth monomethoxide (MOCH$_3$) family. Individual STMs form qubits, and the system is readily scalable to 100 to 1000 qubits. STM qubits have desirable features for quantum computing compared to atoms and diatomic molecules. The additional rotational degree of freedom about the symmetric top axis gives rise to closely-spaced opposite parity $K$-doublets that allow full alignment at low electric fields, and the hyperfine structure naturally provides magnetically insensitive states with switchable electric dipole moments. These features lead to much reduced requirements for electric field control, provide minimal sensitivity to environmental perturbations, and allow for 2-qubit interactions that can be switched on at will. We examine in detail the internal structure of STMs relevant to our proposed platform, taking into account the full effective molecular Hamiltonian including hyperfine interactions, and identify useable STM qubit states. We then examine the effects of the electric dipolar interaction in STMs, which not only guide the designing of high-fidelity gates, but also elucidate the nature of dipolar spin-exchange in STMs. Under realistic experimental parameters,  we estimate that the proposed QC platform could yield gate errors at the $10^{-3}$ level, approaching that required for fault-tolerant quantum computing. 
\end{abstract}

\date{\today}
\maketitle
\section{Introduction}
Universal quantum computing (QC) promises to deliver exponential speed-ups to challenging computational problems -- ranging from large integer factorization \cite{shor1999polynomial} and solving linear systems of equations \cite{harrow2009quantum} to simulation of quantum many-body systems~\cite{lloyd1996universal,wiesner1996simulations,abrams1997simulation}. Many approaches to quantum computation have been explored in the last two decades, including trapped ions and neutral atoms \cite{wang2016single,Lukin_51atom_2017,benhelm2008towards,ballance2016high}, cavity QED and nonlinear optics \cite{casabone2013heralded,reiserer2015cavity,sipahigil2016integrated}, as well as superconducting circuits \cite{neill2018blueprint} and spin-based systems \cite{vandersypen2005nmr,mouradian2015scalable}. Approaches using ultracold polar molecules, in particular, have gained traction in recent years as a potential platform for quantum computing~\cite{demille2002quantum,yelin2006schemes,karra2016prospects,Blackmore2018tweezerQI,Ni2018,Hudson2018}. Ultracold molecules could offer the coherence times of neutral atoms~\cite{Park2017coherence}, and in addition, strong controllable long-range interactions~\cite{demille2002quantum,yelin2006schemes,karra2016prospects}. Futhermore, molecule-based quantum computing platforms offer the prospect of coupling to solid-state devices operating in the microwave regime~\cite{rabl2006hybrid}. These can eventually pave the way towards hybrid systems that couple to photon-based qubits, which are ideally suited for transmission of quantum information.

The many desirable features of molecules arise from their rich internal structure, which has made full quantum control of molecules prohibitively difficult until recently. In the past few years, direct laser cooling and magneto-optical trapping of diatomic molecules (SrF, CaF, and YO) has been demonstrated~\cite{barry2014SrF,McCarron2017magtrap, norrgard2016SrFsubmillikelvin,Anderegg2018ODT,hummon2013YO}, enabling the rapid production of large, trapped samples of ultracold molecules. Beyond diatomic molecules, laser cooling of polyatomic species (SrOH) have been demonstrated~\cite{KozyryevSisyphus,KozyryevBCF}, with ongoing efforts towards magneto-optical trapping of triatomic species. The laser cooling techniques developed so far are readily extendable to more complex polyatomic species, such as the isoelectronic species CaOCH$_3$~\cite{isaev2015polyatomic,KozyryevProposalforLaserCooling}.

\section{Overview of the Proposed Experimental Platform}
In this work, we propose a new platform for quantum computing - an array of optically trapped symmetric-top molecules (STMs). STMs possess $C_{3v}$ symmetry, giving rise to closely-spaced opposite parity $K$-doublets (Fig.~1(a)). These doublets allow for orientation at small electric fields of $\sim 10\,\text{V/cm}$, many orders of magnitude lower than in diatomic molecules. As we will show, more features desirable QC arise naturally from the internal molecular structure of STMs. The idea of using STMs as a quantum resource for generating entangled states was first proposed in~\cite{Wei2011}. Here, we present a specific scheme for quantum computing with STMs and provide estimates of achievable fidelities, taking into account spin-rotation and hyperfine structure and realistic experimental parameters. 

\begin{figure}[h!]
\includegraphics[width=\columnwidth]{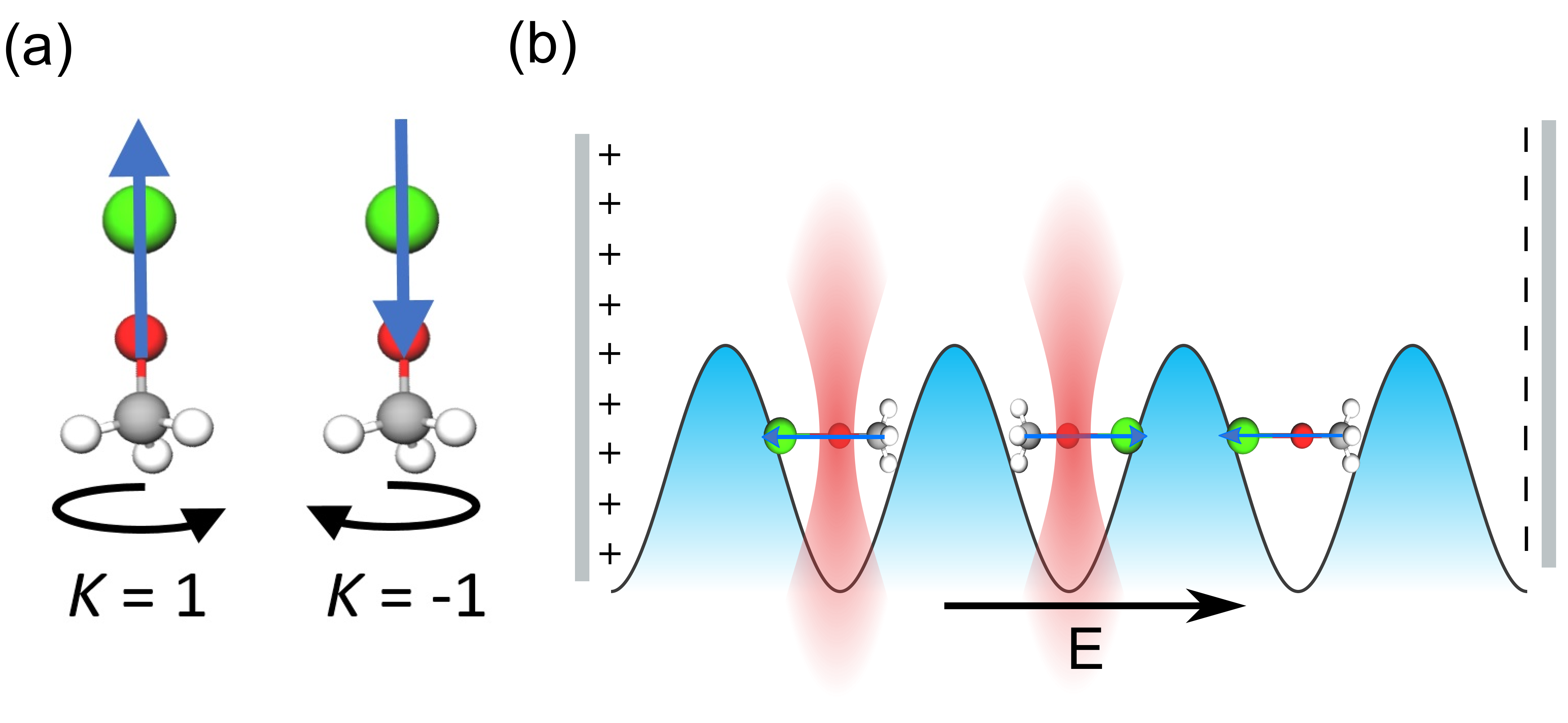}
\caption{(a) The $C_{3v}$ symmetry of STMs produces closely-spaced $K$-doublets, allowing full orientation of the molecules at low DC electric fields.
(b) Proposed platform for a scalable universal quantum computer based on STMs. Ultracold STMs are trapped in a closely-spaced 1D optical lattice (blue). A high-resolution imaging path allows high-fidelity optical readout and state preparation. Tightly-focused laser beams (red) allow addressing of individual STM qubits. }
\end{figure}

Our proposed setup consists of a uniform array of trapped STMs in a closely-spaced optical lattice (Fig.~1(b)), where each lattice site is occupied by a single molecule that forms a single qubit. Tightly-focused laser beams are used to address individual molecules. Combined with microwaves, arbitrary local single-qubit rotations can be performed. To achieve universal quantum computing, we implement a CNOT gate by making use of the dipolar energy shift between neighboring molecules that are polarized. Molecules are prepared in the $|K|=1$ manifold and aligned with a DC electric field along the lattice axis. By changing the internal state of a molecule, the lab frame dipole moment can be switched on and off at will. By an appropriate sequence of pulses discussed below, CNOT gates between pairs of neighboring qubits can be achieved.

There are many advantageous features in our approach. First, it is scalable. With a 1D array, 10s to 100s of uniformly occupied qubit arrays can be created, as demonstrated recently for neutral atoms utilizing optical traps~\cite{Endres2016,Barredo2016}. Future extensions to 2D arrays can reach array sizes ranging from 100s to 1000s of qubits. Second, our approach has fast cycle times. Based on established methods of generating bright slow beams of cold molecules~\cite{Doyle:12} and demonstrated performance of producing trapped diatomic molecules~\cite{McCarron2017magtrap,Williams2018magtrap,Anderegg2018ODT}, large arrays of MOCH$_3$ molecular qubits could be produced in $\sim 100\,\text{ms}$. Third, our approach vastly reduces the technical requirement in electric field control for QC using polar molecules. Fourth, our choice of qubit states leads to much reduced sensitivity to ambient magnetic fields, a major source of decoherence in ultracold atomic systems. 

In the subsequent sections, we will discuss in detail: 1) producing and detecting optically trapped arrays of STM, 2) 2-qubit CNOT gate scheme, 3) choice of qubit states, and 4) estimates of CNOT gate fidelity under realistic experimental conditions.

\section{Producing and Detecting Arrays of Optically Trapped STMs}
Our proposal relies on cooling, trapping, and manipulating STMs optically. This approach requires ultracold temperatures ($\ll 1\,\text{mK}$), as far-detuned optical traps are limited in depth to $\sim 1\,\text{mK}$. An efficient and direct way of producing ultracold matter is via laser cooling, which relies on the scattering of many photons to reduce motional energy. For molecules, this is a major technical challenge since molecules typically have a large number of vibronic decays to dark vibrational states. Nevertheless, in recent years, many molecules~\cite{isaev2015polyatomic}, including MOCH$_3$ molecules~\cite{KozyryevProposalforLaserCooling}, have been identified as favorable for laser-cooling. In fact, in the past few years, laser cooling of diatomic and triatomic molecules has been demonstrated, and large dense samples of ultracold diatomic molecules are now routinely produced via this method.
\begin{figure}
	\includegraphics[width=\columnwidth]{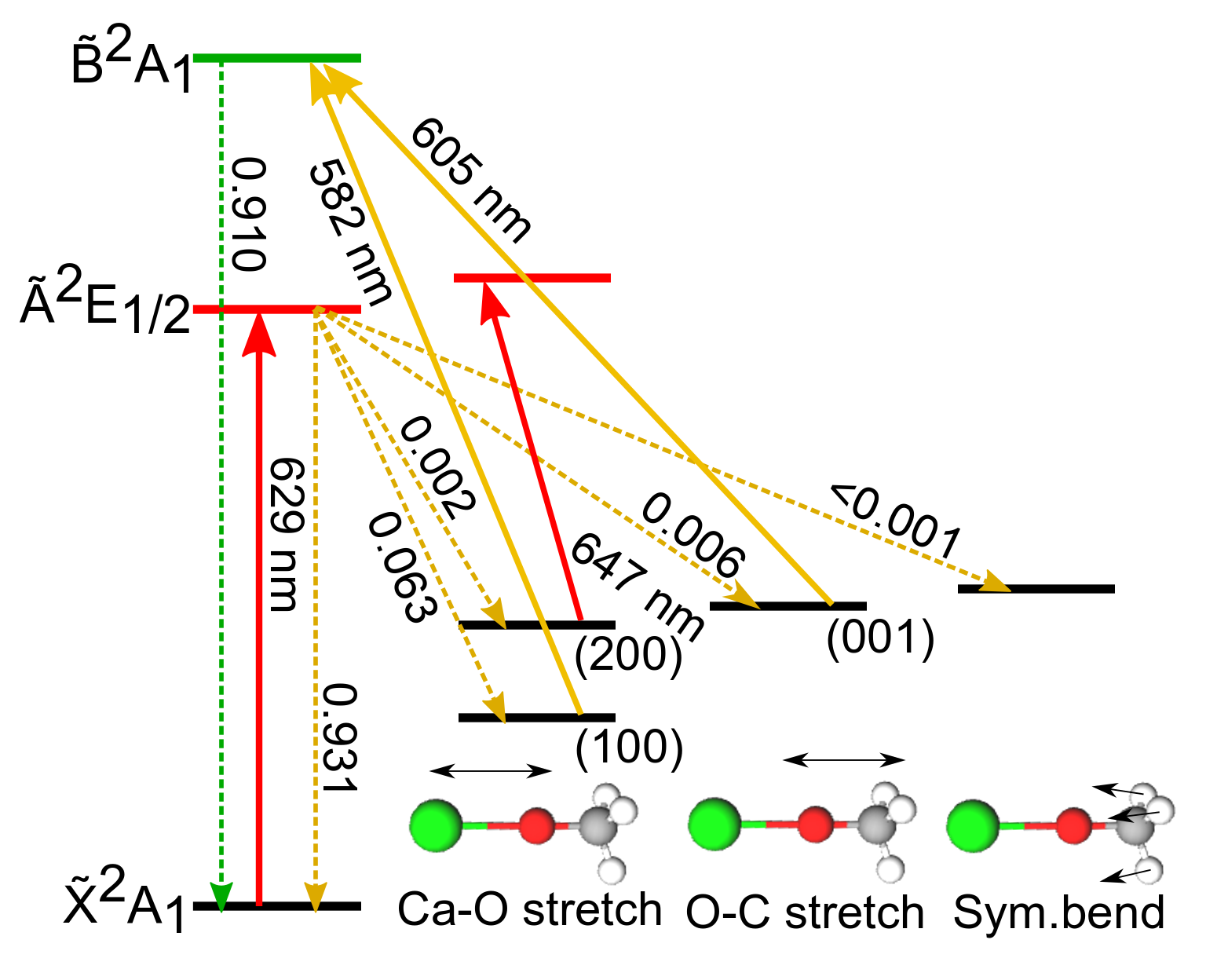}
	\caption{Optical cycling scheme for CaOCH$_3$. Shown are the wavelengths of the transitions, and the vibrational branching ratios~\cite{Kozyryev2019}. The vibrational numbers are labeled in analogy to linear triatomic molecules (ie. CaOH).}
\end{figure}

Recently, a vibrationally-closed photon cycling scheme for the symmetric top molecule CaOCH$_3$ was identified~\cite{Kozyryev2019}, where fewer than $5$ repump lasers (an experimentally manageable task) was needed to scatter $\sim 10^5$ photons. This number is sufficient for creating a magneto-optical trap (MOT), which can subsequently be loaded into a closely-spaced ($\sim 0.5\mu\text{m}$) 1D optical lattice (Fig. 1) using sub-Doppler cooling~\cite{truppe2017CaF,Cheuk2018,Anderegg2018ODT}. Within each lattice site, one expects to reach the collisional blockade regime, where each site has no more than one molecule. Loading of single molecules into optical tweezers with these techniques has been recently demonstrated for CaF molecules~\cite{Anderegg2019Tweezer}. Subsequent rearrangement of this sparsely occupied array, using non-destructive optical detection (recently demonstrated for CaF), along with reprogrammable optical tweezer beams (demonstrated for neutral atoms) will allow creating defect-free arrays of STMs.

In addition to enabling production of trapped samples at ultracold temperatures, the ability to cycle many photons also enables high-fidelity state preparation and readout. At photon detection efficiency of unity, the photon budget of $\sim10^5$ sets a theoretical minimum infidelity of $\sim10^{-5}$. In practice, a photon detection efficiency of $>0.1$ can be achievable with a high numerical aperture microscope objective, giving infidelities of $\sim10^{-4}$. Non-destructive imaging at similar infidelities should also be possible. Non-destructive imaging has in fact been demonstrated recently for optically trapped samples of CaF~\cite{Cheuk2018, Anderegg2019Tweezer}, where detection fidelities exceeding $90\%$ were achieved, limited by low light collection efficiency and vacuum lifetime. An analogous scheme of $\Lambda$-imaging can applied to MOCH$_3$ molecules in the $N=1, |K|=1$ states. With technical improvements in vacuum lifetime to $\sim 10\,\text{s}$, one expects to approach the infidelity limit of $\sim 10^{-4}$ set by the finite photon budget. 

To extend the high-fidelity detection into state-selective detection, one can implement a ``shelving'' technique. Using microwave pulses, molecules can be shelved from qubit states in $N=1$ into other rotational states in $N=2$, which are far off-resonant from the light used for imaging $N=1, |K|=1$. Microwave transitions allow internal states to be spectroscopically resolved, and different hyperfine states can be brought back to the $N=1, |K|=1$ manifold for imaging. Information about the internal state is thus mapped onto the population, which is directly detected. By relying on two-photon processes, the shelving can further be made to be spatially selective. Individual molecules can be addressed. This opens the door to measurements of parts of the system, important in some schemes of error correction, or measurement-based quantum computing.

\section{Single Qubit Gates and 2-Qubit CNOT Gate}
We propose to form qubit states out of the $\ket{N=1, |K|=1}$ manifold of a MOCH$_3$ molecule in the presence of a DC electric field. Transitions between these states are addressed with microwaves, which allow arbitrary global single-qubit rotations. Local rotations are also possible by addressing individual qubits through tightly-focused beams (Fig.~1). For example, qubits can be shifted into resonance by making use of the AC Stark shift of the tightly-focused beams \cite{Weitenberg2011, Wang2015}. Alternatively, 2-photon Raman transitions can be used. Single qubit gate fidelities similar to those in neutral atoms can thus be achieved.

To create two-qubit gates, interactions between two qubits are required. In atom-based approaches, one makes uses of highly excited electronic states~\cite{Branden2010}. These states have short lifetimes ($\sim 100\,\mu s$), which set a fundamental limit on coherence time. In addition, the AC polarizabilities of highly-excited electronic states can often be very different from ground electronic states, rendering simultaneous trapping difficult. The key advantage of molecule-based approaches is the availability of long-range interactions \textit{without} using highly-excited electronic states. We estimate that the proposed states used both for single-qubit rotations and 2-qubit gates will have lifetimes of ($\sim 10\,\text{s}$), limited only by black-body excitation to vibrationally excited states. Coherence time on the second scale has in fact been demonstrated for optically trapped bi-alkali molecules~\cite{Park2017coherence}, and we expect that this would also be possible for STMs in the future.

To implement 2-qubit gates in molecules, various schemes have been proposed~\cite{demille2002quantum, Wall2009, Ni2018, Hudson2018}. Here we propose to create a 2-qubit CNOT gate by relying on the dipolar blockade~\cite{Lukin2001}, which has been demonstrated using neutral Rydberg atoms~\cite{isenhower2010demonstration}. A DC electric field aligns the STMs, producing internal states that have different lab frame dipole moment. Crucially, states with zero dipole moments are available. We pick two zero dipole moment states as qubit states $\ket{0}$ and $\ket{1}$. In addition, we use a third auxiliary state $\ket{e}$ that has a large electric dipole moment. $\ket{0}$ and $\ket{1}$ are non-interacting, which make them ideal for storing quantum informaiton. Molecules in $\ket{e}$ interact via the electric dipolar interaction, resulting in an energy shift that can be used to create a 2-qubit gate. Thus by simply switching between $\{\ket{0},\ket{1}\}$ and $\ket{e}$, 2-qubit gates can be turned on and off on demand.

\begin{figure}
	\includegraphics[width=\columnwidth]{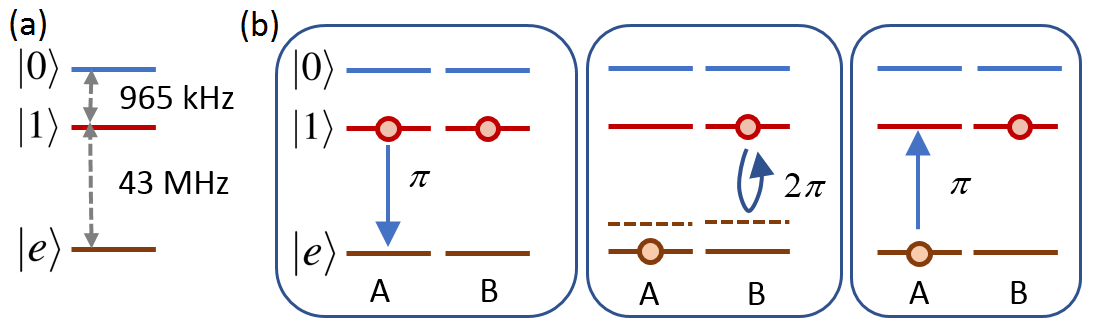}
	\caption{(a) Energy spacing of the qubit states $\ket{0}$, $\ket{1}$ and $\ket{e}$. (b) Pulse sequence used for implementing a $C_z$ gate. Qubits A and B are the control and target qubits respectively. Shown is the case when both qubits are in $\ket{1}$. The dipolar blockade shifts the $2\pi$ pulse on qubit B out of resonance. The three pulses together give an overall $\pi$ phase shift of the 2-qubit state.}
	\label{fig:blockade}
\end{figure}

For implementing a 2-qubit CNOT gate, we specifically consider the following scheme (Fig.~\ref{fig:blockade}). Two qubits out of the molecular array act as the control qubit and the target qubit respectively. A $C_z$ gate is implemented with a standard sequence of pulses shown in Fig.~3(b), consisting of a single $2\pi$ pulse on the target qubit applied between two $\pi$ pulses on the control qubit on the $\ket{1}\rightarrow \ket{e}$ transition. Because of the dipolar blockade mechanism when both qubits are in $\ket{1}$, the pulse sequence results in a phase shift of $\approx \pi$ unless both qubits are in $\ket{0}$. To create a CNOT gate from the $C_z$ gate, $\pi/2$-pulses on the target qubit are applied before and the $C_z$ step.

\section{Dipolar Interactions in STMs and Choice of Qubit States}
\begin{figure}[h!]
	\includegraphics[width=0.9\columnwidth]{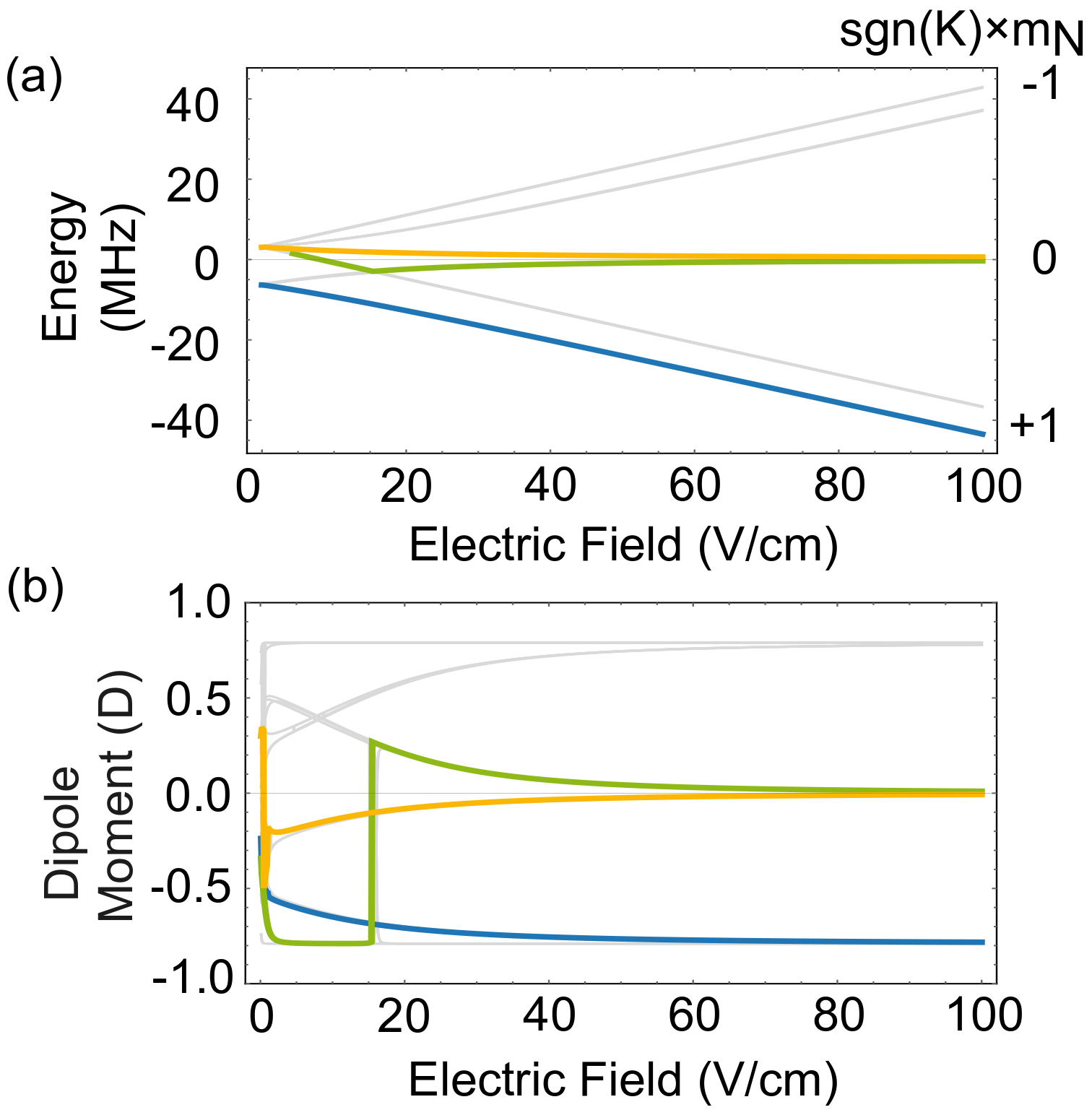}
	\caption{(a) Energy of $N=1, |K|=1$ states versus electric field for CaOCH$_{3}$. (b) Lab frame dipole moments of $N=1, |K|=1$ states versus electric field for CaOCH$_{3}$. For both plots, the qubit states $\ket{0}$, $\ket{1}$, and $\ket{e}$ are shown in solid (green, yellow, blue). At high electric fields, the states are split into three manifolds by $\text{sgn}(K)m_N$. 
	}
	\label{fig:levels}
\end{figure}

A key feature in STMs are closely spaced opposite parity $K$-doublets (for $|K|>0$), which arise from the rotational degree of freedom about the symmetric top axis. In particular, $|K|=1, N=1$ states enjoy a large range in electric fields where Stark shifts are linear, or equivalently, the electric dipole moments along the quantization axis are constant. For this manifold, in the linear Stark shifts regime, states are split into 3 groups with positive, negative, and zero lab frame electric dipole moments, corresponding to $\text{sgn}(K)\times m_N=-1,0$, or $1$, where $m_N$ is the projection of the angular momentum excluding spin along the quantization axis (See Fig.~\ref{fig:levels}). Note that in this ``high-field'' regime, $m_N$ is an approximately good quantum number. The manifold of $m_N=0$ states is crucial in our scheme, as they provide ideal qubit states minimally sensitive to dipolar interactions. In addition, constant dipole moments give the practical advantage that the strength of dipolar interactions is minimally sensitive to the precise strength of the electric field, which reduces the demands of field control.

We additionally desire qubit states to be spectroscopically resolvable. This presents a challenge in STMs, since the additional degrees of freedom leads to closely-spaced levels whose properties require a detailed understanding of the internal structure, including hyperfine effects. We have examined in detail the internal state structure for the molecule CaOCH$_3$, and find that resolvable states with the desired properties of switchable dipole moment and minimal magnetic field dependence are available.
More generally, we find that such states are always available in the $N=1, |K|=1$ STMs of Hund's case (b) in the high electric field regime, even in the absence of a magnetic field.

In detail, in a Hund's case (b) STM, the internal states can be labeled by $\ket{N,K,S,J,I, F, m_F}$, where $N$ is the angular momentum excluding spin, $K$ is its projection along the molecular axis, $S=1/2$ is the electron spin, $J=N+S$ is the total angular momentum excluding nuclear spin, $I$ is the nuclear spin, $F=J+I$ is the total angular momentum and $m_F$ is its projection. We focus on states with $\ket{N=1,|K|=1,S=1/2,J,I, F, m_F}$ in the high-field limit, where $I=1/2$, and $m_N$ is an approximately good quantum number (see Supp. Mat. on value of $I$). Without additional coupling, each $\text{sgn}(K) \times m_N$ manifold is 8-fold degenerate (2-fold in $S$, $I$ and $\text{sgn}(K)$ ). Although the spin-rotation and the isotropic hyperfine interaction lifts the degeneracies in $m_S$ and $m_I$, the degeneracy in $\text{sgn}(K)$ remains. The anisotropic hyperfine interaction in STMs can couple states that have $\Delta K=2$. However, this coupling is weak, and is only leads to spectroscopically resolvable splittings when the states are near-degenerate. 

To identify states that are coupled, we note that generically, $\ket{N,K,S,J,I, F, m_F}$ and  $\ket{N,-K,S,J,I, F, -m_F}$ are degenerate in the absence of the $\Delta K=2$ interaction. Since this interaction preserves the total lab frame angular projection $m_F$ (See Supp. Mat.), it only strongly couples states with $m_F=0$ for $|K|=1$ states, since they remain degenerate upon changing $K \rightarrow -K$. Since $m_F$ remains a good quantum number in the presence of electric and magnetic fields along the quantization axis, this condition alone identifies resolvable states in the linear Stark shift regime: States where $m_N+m_S+m_I=0$ form $K$-doublets. It is clear then that the $\text{sgn}(K)\times m_N=\pm 1$ manifolds each have a pair of states ($m_s=m_I=-m_N/2, K=\pm1$) that are split by the $\Delta K=2$ hyperfine interaction. Each of these manifolds, which have a non-zero dipole moment, has 2 resolvable states. For the $K\times m_N=0$ manifold, which has vanishing lab frame dipole moment, has two pairs of states ($m_s=-m_I=\pm1/2, K=\pm1$) split by the $\Delta K=2$ hyperfine coupling, giving a total of 4 resolvable states. 

\begin{figure}[h!]
	\includegraphics[width=0.9\columnwidth]{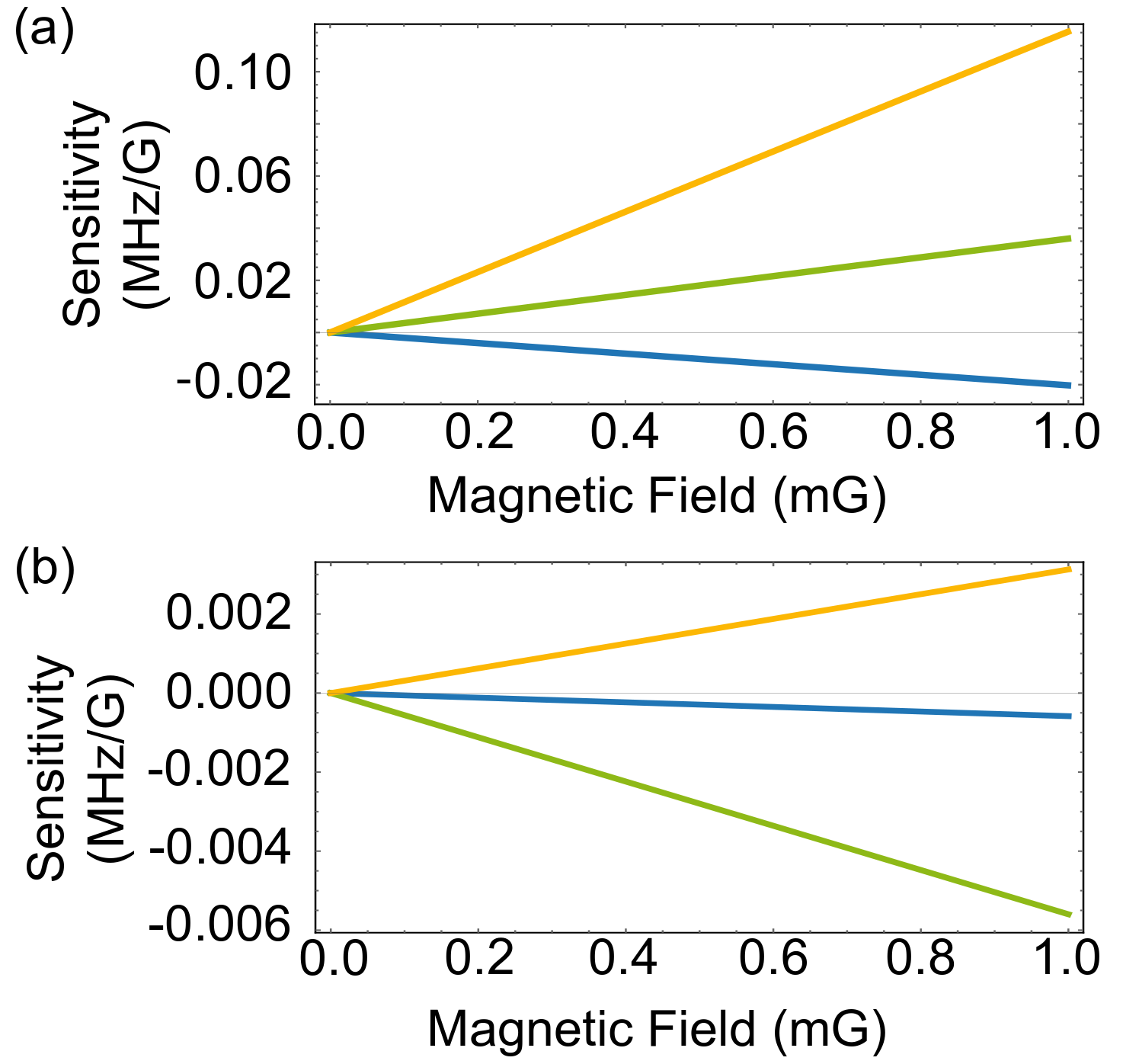}
	\caption{(a) Magnetic field sensitivity of qubit states versus magnetic field strength along the quantization axis for CaOCH$_{3}$ at $100\,\text{V/cm}$. 
	(b) Magnetic field sensitivity of qubit states versus magnetic field strength orthogonal to the quantization axis for CaOCH$_{3}$ at $100\,\text{V/cm}$. For both plots, green, yellow, and blue solid lines correspond to the states $\ket{0}$, $\ket{1}$, and $\ket{e}$. The magnetic sensitivities are linear in field and approximately vanish at zero applied field, indicating a quadratic dependence on magnetic field. Note that at a field of $1\,\text{mG}$, the magnetic sensitivities of these states are much less than that corresponding to the Bohr magneton $\mu_B$ (1.4\,MHz/G).
	}
	\label{fig:magnetic}
\end{figure}

While the energies of the resolvable states depend on specific values of each STM molecule and the value of the external field, out of the 24 $N=1, |K|=1$ states, there are always 4 resolvable states with zero dipole moment and 2 resolvable states for each sign of the dipole moment. In addition, these resolvable states have zero electron spin projection, since they have equal admixtures of states with $\left\{K,m_S=\frac{1}{2}\right\}$ and $\left\{-K,m_S=-\frac{1}{2}\right\}$. Ignoring the contribution from the nuclear spin (much smaller than that from the electron spin), these $K$-doublet states thus have vanishing magnetic field sensitivity along the quantization axis. They also have vanishing sensitivity orthogonal to the quantization axis, since each component separately has zero spin projection orthogonal to the quantization axis, and no interference between two components occurs since they differ in the quantum number $K$. Note that the same argument applies to $m_I$, which implies that at zero field, the magnetic moments of these $K$-doublets are zero. The vanishing sensitivity to magnetic fields is shown in Fig.~\ref{fig:magnetic} specifically for 3 of these states in CaOCH$_3$ at a field of 100\,V/cm.

Any 2 of the 4 resolvable zero dipole moment states ($m_N=0$) are therefore ideal qubit storage states $\ket{0}$ and $\ket{1}$. They are insensitive to electric field fluctuations owing to zero lab frame dipole moment, and insensitive to ambient magnetic field fluctuations, which is a major source of decoherence in experiments. Any of the 4 resolvable states in $m_N=\pm 1$ can be used to couple two molecules to create 2-qubit gates. 

To further guide the selection of qubit states, we have to examine the dipolar interactions of these resolvable states that have a non-zero dipole moment. The dipolar interaction between two molecules is given by
\begin{equation}
H_{dd}=\frac{1}{4\pi \epsilon_0 r^3}\left[\hat{\mathbf{d}}_1\cdot \hat{\mathbf{d}}_2-3(\hat{\mathbf{d}}_1\cdot \hat{\mathbf{r}})(\hat{\mathbf{d}}_2\cdot \hat{\mathbf{r}})\right],
\label{eqn:Hdd}
\end{equation}
where $\mathbf{r}$ is the intermolecular separation. For the specific case we propose, where the electric field is along the inter-molecular axis, the interaction takes the form 
\begin{eqnarray}
H_{dd,0}&=&\frac{-1}{2\pi \epsilon_0 r^3}\left[T^1_0(\mathbf{d}_1)T^1_0(\mathbf{d}_2) \right. \nonumber \\
& & \left. +\frac{1}{2}\left(T^1_1(\mathbf{d}_1)T^1_{-1}(\mathbf{d}_2)+T^1_{-1}(\mathbf{d}_1)T^1_1(\mathbf{d}_2)\right) \right],
\label{eqn:Hdd0}
\end{eqnarray}
where $T^1_p(\mathbf{V})$ denote spherical tensor components of a rank-1 operator $\mathbf{V}$ with quantization axis along the electric field. 

In this form, two effects are apparent. First, there is an energy shift when the lab frame dipole moments of both molecules are non-zero. First-order perturbation theory gives 
\begin{equation}
\langle i_1 i_2 | H_{dd,0} | i_1 i_2\rangle \propto \langle i_1| T^1_0(\mathbf{d}_1) |i_1 \rangle \langle i_2|T^1_0(\mathbf{d}_2)| i_2 \rangle,
\end{equation}
where $i_1$ and $i_2$ denotes the internal state of the two molecules. This shows that the energy shift is proportional to the product of the lab-frame dipole moments. Note that the second term in Eq.~(\ref{eqn:Hdd0}) does not contribute, since it must change the total angular momentum projection $m_F$ of each molecule and hence its state. It however still conserves the sum of the total angular momentum projection of both molecules along the quantization axis. For $N=1,|K|=1$ STM states, we note that the dipolar energy shift is determined by which $\text{sgn}(K)m_N$ blocks the states are in, i.e. by the lab frame dipole moments along the quantization axis.

Second, in addition to an energy shift, dipolar spin-exchange is also present. Two molecules in states $i_1$ and $i_2$ can be in two degenerate configurations $\ket{i_1,i_2}$ and $\ket{i_2,i_1}$. In this process, all terms in $H_{dd,0}$ can contribute. The matrix element is dependent on the terms of the form $\langle i_2 i_1| T^1_p(\mathbf{d}_1)T^1_{-p}(\mathbf{d}_2)|i_1 i_2\rangle$. For a given pair of internal states $i_1$ and $i_2$, it suffices to evaluate $\langle i_2|T^1_p(\mathbf{d})|i_1\rangle$, that is, the transitional dipole moment between the two states. Physically, this shows that the spin-exchange process can be thought of as an electric dipole transition on molecule 1 driven by the electric field arising from the electric dipole of molecule 2.

In the proposed scheme of implementing a CNOT gate via the dipolar blockade, we want to maximize the energy shift, but avoid any spin-exchange effects. Spin-exchange leads to bit flip errors and effective qubit loss due to population leakage outside $\{\ket{0}, \ket{1}\}$. Although one generically expects the spin-exchange coupling to be comparable to the dipolar energy shift, since they both arise from the same underlying dipolar interaction, spin-exchange can in fact be suppressed by a careful choice of states. As mentioned above, in the high-field regime, the electron spin projection $\langle S_z\rangle$ for resolvable states vanishes. This is because the resolvable states, which are $K$-doublets, can be written approximately as $\frac{1}{\sqrt{2}}\left(|m_S=\frac{1}{2}, K, m_N,\eta\rangle \pm  |m_S=-\frac{1}{2}, -K,-m_N,\eta' \rangle\right)$, where $\eta$ and $\eta'$ denote omitted quantum numbers. We will henceforth call the relative sign between these two dominant components the spin-parity of the state. It is an approximately good quantum number. 

Next, we note that the dipole moment operator $T^1_p(\mathbf{d})$ does not act on the electron spin or $K$, implying that the spin-parity is approximately conserved. Thus, states with identical spin-parity experience suppressed spin-exchange rates. We therefore pick the qubit states $\ket{0},\ket{1},\ket{e}$ to be of the same spin-parity. For CaOCH$_3$, at an electric field of $100\,\text{V/cm}$, this can be accomplished by identifying $\ket{0}=\ket{i=12}$, $\ket{1}=\ket{i=16}$ and $\ket{e}=\ket{i=1}$, where $i$ denotes that $i$th eigenstate in the $N=1, |K|=1$ manifold when ordered by energy from lowest to highest. This is the choice of qubit states used for subsequent calculations. The suppression of spin-exchange with this choice of states is verified numerically. An alternative scheme using opposite spin parities is $\ket{0}=\ket{i=11}$, $\ket{1}=\ket{i=15}$ and $\ket{e}=\ket{i=2}$. This scheme has slightly higher magnetic field sensitivity away from zero magnetic field and is therefore not used.

Our choice of qubit states to be of the same spin-parity suppresses spin-exchange, since transitions between these states are M1 transitions. This however presents a new problem. Generally, M1 transitions are much weaker than E1 transitions, with Rabi frequencies suppressed by $c d/\mu$, where $c$ is the speed of light, $d$ the dipole moment, and $\mu$ the magnetic moment. With $d=1\,\text{D}$ and $\mu= 1\mu_B$, this factor is $\sim100$. Consequently, when one drives M1 transitions with an oscillating electromagnetic field, off-resonant E1 transitions could be non-negligible due to their much stronger strengths. This can be problematic for STMs, since transitions can have frequency differences as small as 10s of kHz. For example, the frequency difference between hyperfine states can be as low as 50\,kHz in CaOCH$_3$ at $100\,\text{V/cm}$, which implies that the Rabi frequency for M1 transitions would have to be much less than 500\,Hz in order to avoid significant off-resonant excitation. To avoid off-resonant E1 transitions, one can rely on two-photon Raman transitions where both photons drive strong E1-allowed transitions to the excited electronic B state. This has the further benefit of allowing individual addressing of each qubit. With the two-photon process driving effective M1 transitions between the qubit states, the Rabi frequency to states directly connected via E1 transitions in the $N=1, |K|=1$ manifold are now suppressed by $c d/\mu\sim 100$. They are thus suppressed both by being off-resonant, and by the approximate selection rule.

\section{CNOT Gate Fidelity}
In this section, we estimate the gate fidelity of the 2-qubit CNOT gate in the proposed scheme. We use the molecule CaOCH$_3$ at an electric field of $100\,\text{V/cm}$. The inter-molecule spacing is set to be $0.5\,\mu\text{m}$.

We first estimate intrinsic errors for a 2-qubit CNOT gate using a dipolar blockade scheme. This scheme has been previously used in Rydberg atoms, and the intrinsic error of the gate induced by incomplete blockade in the limit of infinite lifetime is given by~\cite{saffman2005analysis,saffman2010quantum}: 
\begin{equation}
	\varepsilon=\frac{\Omega^2}{8U^2_{dd}/\hbar^2}\bigg(1+\frac{6\Omega^2}{\omega_{10}^2}\bigg),
	\label{eq:analyticerror}
\end{equation}
where $\Omega$ is the Rabi frequency of the $\ket{1}\to\ket{e}$ coupling, $\omega_{10}$ is the frequency splitting between the $\ket{1}$ and $\ket{0}$ states, and $U_{dd}$ is the dipole blockade energy. The infinite lifetime approximation is a good approximation in the absence of technical imperfections, since $\ket{e}$ states have blackbody limited lifetimes $\sim 10\,\text{s}$, much longer than the expected single and 2-qubit gate times of $\sim 1\,\text{ms}$. In our case, $U_{dd}/\hbar=2\pi\times 1.477\,\text{kHz}$ for a molecule spacing of $0.5\,\mu$m and an applied field of $100\,\text{V/cm}$. For $\Omega=2\pi\times 50\,\text{Hz}$, Eq.~(\ref{eq:analyticerror}) estimates an error of $1.4\times10^{-4}$.

The intrinsic error rate could be decreased arbitrarily at the cost of longer gate time. Nevertheless, in the presence of decoherence and spin-exchange, an optimal gate speed will be obtained at finite duration. 
For example, even spin-exchange among the qubit states $\left\{\ket{0},\ket{1},\ket{e}\right\}$ can lead to significant errors, as the molecules can be lost from the logical basis $\left\{\ket{0},\ket{1}\right\}$. Off-resonant coupling could transfer a qubit into states outside $\left\{\ket{0},\ket{1},\ket{e}\right\}$, resulting in effective loss of the qubit.

To estimate the intrinsic errors of our scheme, we first assume that single qubit rotations have negligible intrinsic errors. This is a valid assumption, since the Rabi frequency driving single qubit rotations can be made much faster than the estimated decoherence rates of $1\,\text{s}^{-1}$ (see Supp. Mat.). It therefore suffices to examine the $C_z$ gate alone. As shown in Fig.~\ref{fig:blockade}, the $C_z$ gate consists of 3 parts, two $\ket{1}\rightarrow\ket{e}$ $\pi$ pulses with a $\ket{1}\rightarrow\ket{e}$ $2\pi$ pulse that resolves the dipolar blockade energy shift in between. We characterize the $\pi$ pulse and the $2\pi$ pulse separately.

To characterize the $\pi$ pulse, we numerically integrate Schrodinger's equation with the $24\otimes 24$ 2-qubit Hamiltonian containing all $N=1, |K|=1$ states and a time-dependent drive. The initial state is $\ket{1}\otimes\ket{1}$. All possible M1 transitions are considered for the time-dependent drive, and we do not make the usual rotating wave approximation. Given the spacing of hyperfine states, the rotating and counter-rotating contributions can be similar for off-resonant excitations. Since E1 transitions will be suppressed by $\sim100$ in a two-photon scheme discussed in the previous section, we do not take these into account. Using an effective M1 coupling with $\ket{1}\rightarrow\ket{e}$ Rabi frequency of $\Omega = 2\pi \times 16\,\text{kHz}$ and polarization along the quantization axis, we find that $\pi$-pulses between $\ket{1}\to\ket{e}$ can be accomplished with an error of $3\times 10^{-6}$. Here, the error is defined as the deviation of the squared overlap between the ideal and simulated states from unity, $1-|\braket{\psi_\text{sim}|\psi_\text{ideal}}|^2$. Note that since we do not have to resolve the blockade shift in this step, a Rabi frequency much larger than the blockade shift $U_{dd}$ can be used. 

\begin{figure}[h!]
\includegraphics[width=\columnwidth]{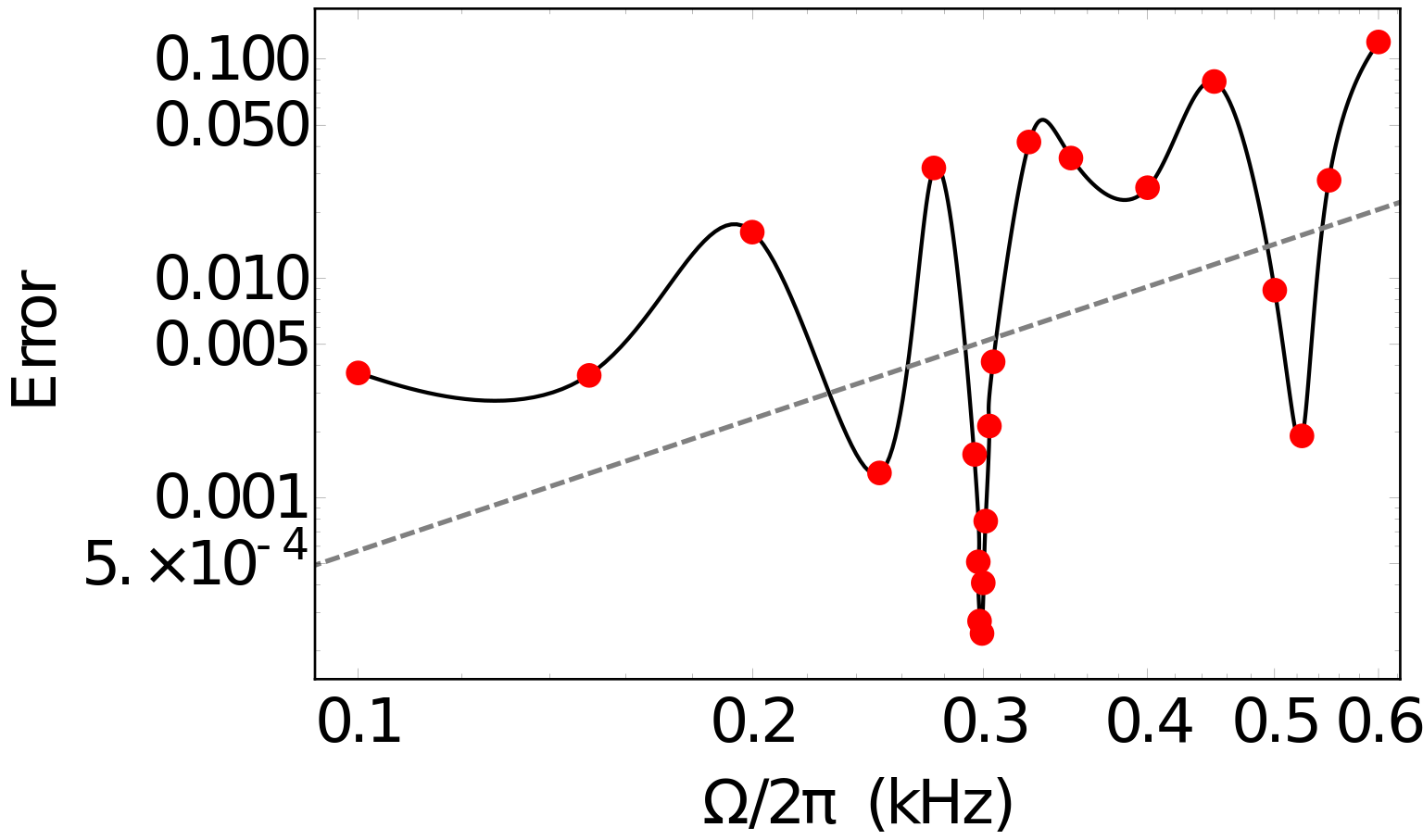}
\caption{The error of the $\ket{1}\rightarrow\ket{e}$ $2\pi$ pulse versus Rabi frequency $\Omega$. Dashed line shows the analytic estimate given by Eq.~\ref{eq:analyticerror}. The numerical results show that fine-tuning of the Rabi frequency can lead to errors lower than Eq.~\ref{eq:analyticerror}. }
\label{fig:rabi}
\end{figure}

We next characterize the $2\pi$ pulse, where the dipole blockade energy has to be resolved. We again numerically integrate Schrodinger's equation with the $24\otimes 24$ 2-qubit Hamiltonian and a time-dependent drive with the initial state $\ket{1}\otimes\ket{1}$. With a fine-tuned Rabi frequency of $\Omega=2\pi\times 300\,\text{Hz}$, the average error is found to be $5\times10^{-4}$. The error of the $2\pi$ pulse versus Rabi frequency is shown in Fig.~\ref{fig:rabi}. Given the two orders of magnitude between the $2\pi$ pulse and $\pi$ pulse, the dominant intrinsic error of the proposed $C_z$ gate comes from the intermediate $2\pi$ step. Since single qubit rotations are expected to have negligible intrinsic errors, the intrinsic errors of the proposed CNOT gate will be dominated by the errors of the $2\pi$ pulse. 

We next estimate errors from realistic experimental conditions by separately performing Monte Carlo simulations with randomly sampled technical imperfections of both single qubit rotations and the two-qubit CNOT gate operation using a master equation approach with a reduced Hilbert space containing only $\ket{0}$, $\ket{1}$ and $\ket{e}$, along with a rotating wave approximation (RWA). Here, the qubit operations are modeled to be perfect, and experimental errors appear in the master equation in two ways. First, they directly affect the Hamiltonian by modifying detunings and Rabi frequencies. Second, they can lead to dephasing terms captued by the master equation. In detail, microwave and laser power fluctuations lead to Rabi frequency fluctuations and variations in AC Stark shifts. Instabilities in the applied static electric field, as well as fluctuations in the electric and magnetic fields, modify the detuning for the microwave pulses. We have computed the magnetic moments and electric dipole moments of these states taking into account spin-rotation coupling and hyperfine terms (see Supp. Mat.).  A full discussion of field fluctuations including ambient fields and patch potentials are given in the Supplemental Materials. We also partially take into account thermal motion of the molecules. The resulting variations in the interaction strength $U_{dd}$ is considered. Not taken into account is motional decoherence due to state-dependent AC Stark shifts. We defer this to future work on minimizing differential AC Stark shifts. With multi-chromatic repulsive optical traps and a judicious choice of polarizations, ``magic'' conditions where all three states $\ket{0},\ket{1}$ and $\ket{e}$ experience identical trapping potentials are possible, in analogy to similar ideas for bialkali molecules~\cite{Kotochigova2010magic,Rosenband2018magic}. A full list of technical imperfections considered along with the parameters used is shown in Table~\ref{tab:Errors}.  

To isolate the errors that are technical and arise from imperfect control, the master equation is run with Rabi frequency $\Omega=2\pi \times 10\,\text{Hz}$, where blockade errors are well below $10^{-4}$ according to the estimates of Eq.~\ref{eq:analyticerror}. The simulation is run over 10,000 iterations for each input initialization~\cite{johansson2012qutip}. Similar to an analysis previously performed for a Rydberg atomic system~\cite{SaffmanFidelity}, our simulations incorporate the effects of technical imperfections as well as dephasing terms that arise from variations in the local environment of each qubit. We find an average fidelity of the two-qubit CNOT gate under these conditions to be $0.998$. Since dephasing increases with longer gate times, this sets an upper bound to the technical errors for the CNOT gate operated at higher $\Omega$. We thus find that the errors of the proposed 2-qubit CNOT gate will likely be limited by both intrinsic errors and technical errors at the $10^{-3}$ level under currently achievable experimental conditions. At the fine-tuned Rabi frequency, the intrinsic error is limited by technical errors. Future technological improvements will allow higher dipolar interaction energies when molecules are placed closer, potentially allowing intrinsic errors below $10^{-4}$ at the same gate speed (gate time of $\sim 3\,\text{ms}$). Improvements in experimental control will similarly allow operating at slower gate speeds, which in turn will yield higher fidelities. This is promising as these fidelities approach the threshold for fault-tolerant quantum computing.

\begin{table}
\begin{tabular}{|l|l|l|}
\hline
Description  & Value \\
\hline
Molecule Spacing & 0.5\,$\mu$m\\
Applied E-Field &  100\,V/cm\\
Blockade shift  & $h\times$1477\,Hz\\
Applied E-field relative instability &  $10^{-7}$ \\
B-field instability &  $10\,\mu\text{G}$ \\
Phase Error  from E-field &   \\
 patch potentials (see Supp. Mat.) &  $10^{-6}$ \\
Molecular Temperature &   $1$\,$\mu$K\\
Rabi frequency Relative Noise &  $10^{-2}$\\
$\Omega_{\ket{1}\to\ket{e}}$ ($\pi$-pulse Rabi frequency) & $2\pi \times 16\,\text{kHz}$\\
$\Omega_{\ket{1}\to\ket{e}}$ ($2\pi$-pulse Rabi frequency) & $2\pi \times 299\,\text{Hz}$\\
$\omega_{01}$ & $2\pi \times 965$\,kHz\\
$\omega_{1e}$ & $2\pi \times 43.156$\,MHz\\
\hline
\end{tabular}
\caption{Parameters used for numeric simulations of gate fidelities. A full discussion of the used values are provided in the Supplemental Materials. }\label{tab:Errors}
\end{table}

\section{Conclusion}
We have proposed and examined under realistic conditions a platform for universal quantum computing based on optically trapped single STMs. We have also identified new features in STMs that are desirable for creating qubits. For example, much smaller electric fields are needed to align STMs, and dipolar energy shifts can be turned on and off simply by changing internal states. We have also discussed how $K$-doublets that result from the hyperfine interaction are spectroscopically resolvable and magnetically insensitive in the high-field regime. They further give rise to the approximate spin-parity quantum number, which can be used to control dipolar spin-exchange. These considerations make STMs favorable candidates for quantum simulation in addition to quantum computing. In addition to the convenient features of STMs, our proposed platform is also immediately scalable to 10s to 100s of qubits, similar to quantum computing platforms based on optically trapped neutral atoms. 
To obtain estimates of gate fidelities, we have considered major sources of decoherence, the full molecular hyperfine structure and all dipolar interactions effects, and have estimated that the intrinsic CNOT gate errors of our proposed platform can be $<10^{-3}$. We find that under realistic experimental conditions, both technical errors and intrinsic gate errors give errors around $10^{-3}$, while fine-tuning allows intrinsic gate errors to be negligible compared to technical errors. With future technological improvements, errors at the $10^{-4}$ level are possibly within reach. These results are encouraging and suggest that optically trapped STMs could be a viable new platform for quantum computing.

\section*{Acknowledgements}
The authors thank Z. Lasner and B. Augenbraun for fruitful discussions. LWC acknowledges support from the MPHQ. Computations in this paper were run on the Odyssey cluster supported by the FAS Division of Science, Research Computing Group at Harvard University.

\bibliography{ref}
\clearpage

\newpage

\begin{center}
\textbf{\large Supplemental Materials}
\end{center}

\setcounter{section}{0}
\setcounter{equation}{0}
\setcounter{figure}{0}
\setcounter{table}{0}
\makeatletter
\renewcommand{\theequation}{S\arabic{equation}}
\renewcommand{\thefigure}{S\arabic{figure}}

\section{Symmetry of Molecular States}
The Hamiltonian for a MOCH$_3$ molecule can be written as $H_{R}+H_{SR}+H_{CD}+H_{hf}$, where the terms correspond to rotational terms, spin-rotation terms, centrifugal distortion terms, and hyperfine terms.

For a prolate symmetric-top molecule, 
\begin{equation}
	H_{R}= B\hat{N}^2+(A-B)\hat{N}^2_{z}
\end{equation}
where $A$ and $B$ are the symmetric-top rotational constants, $\mathbf{J}$ is the total angular momentum excluding nuclear spin, $\mathbf{N}$ is the total angular momentum excluding spin and $\hat{N}_z$ its projection onto the molecular axis~\cite{hirota}. The eigenenergies of $H_{R}$ are thus given by $BN(N+1)+(A-B)K^2$, where the quantum numbers are given by $N= \langle \hat{N}\rangle$ and $K= \langle \hat{K}\rangle$. Each level has degeneracy of $2\times(2N+1)$, corresponding to the $2N+1$ angular momentum projections in the lab frame, along with the 2 projections of $|K|$ on the molecular axis. 

$H_{SR}$ describes spin-rotation coupling, and is parametrized as
\begin{equation}
	\epsilon_{aa} \mathbf{N}_z \mathbf{S}_z + \frac{1}{2} \epsilon_{bc} (\mathbf{N}_+\mathbf{S}_-+\mathbf{N}_-\mathbf{S}_+)
\end{equation}
The total angular momentum is denoted by $\hat{J} = \hat{N} + \hat{S}$. 

$H_{CD}$ describes centrifugal distortion, and is diagonal in $\hat{N}_z^2$ and $\hat{N}^2$. Since for a fixed $N$ and $K$, $H_{CD}$ only leads to an overall energy shift, it is omitted for subsequent calculations.
$H_{hf}$ describes the hyperfine interactions between the nuclear spins of the H atoms and the electronic spin $S$. We restrict to the case of $|K|=1$.

The symmetric top states $|J,K=\pm 1,M_J\rangle$ transform like $E_{\pm}$. Since the total symmetry of the wavefunction (including nuclear spin) transform like $A_1$, for $|K|=1$, the states allowed by symmetry are
\begin{equation}
\frac{1}{\sqrt{2}}( |K=+1\rangle \otimes |E_-, I\ket \pm  |K=-1\rangle \otimes |E_+, I\rangle  )
\end{equation}
where the symmetry of the nuclear spin function is denoted by $\Gamma$. For three H atoms with spin $1/2$, the nuclear spin states with total $I=3/2$ are clearly symmetric and transform as $A_1$. The two states that have $E_\pm$ symmetry have total nuclear spin $I=1/2$. Thus for this work, where we only consider MOCH$_3$ molecules in the $N=1, |K|=1$ manifold, $I=1/2$.

\section{Anisotropic Hyperfine Coupling}
Here, we give details on the hyperfine coupling that couples $K=\pm 1$ levels. The spin-spin and dipole-dipole interactions between $I$ and $S$ result in coupling between states with $\Delta K\leq 2$.
The matrix elements between basis states with $|K|=1$ is given by
\begin{widetext}
\begin{eqnarray}
& & \langle N',K', S, J', I, \Gamma', F, m_F| \hat{H}_{hf} | N, K, S, J, I,\Gamma F, m_F\rangle  \nonumber \\
&=& \bigg[a_F \delta_{m_F' m_F} \delta_{F' F} \delta_{N' N} \delta_{K' K} (-1)^{N+S'+J'} (-1)^{J'+I +F'+1} \sqrt{(2J'+1)(2J+1)}\nonumber \\
& &\times \sqrt{S(S+1)I(I+1)(2I+1)}\left\{\begin{array}{ccc} I & J' & F \\ J & I & 1 \end{array}\right\} \left\{\begin{array}{ccc} S & J' & N \\ J & S & 1 \end{array}\right\}\bigg] \nonumber \\
& & - \bigg[\delta_{m_F' m_F} \delta_{F'F} (-1)^{J'+I+F'} \sqrt{30} \sqrt{S(S+1)(2S+1)(2J'+1)(2J+1)(2N'+1)(2N+1)} \left\{\begin{array}{ccc} I & J' & F \\ J & I & 1 \end{array}\right\} \nonumber \\
& &\times \left\{\begin{array}{ccc} N' & N & 2 \\ S & S & 1 \\ J' & J &1 \end{array}\right\} 
\sum_{\alpha=0,\pm 1} \langle I \Gamma'|| I_\alpha || \Gamma I_0\rangle \sum_q (-1)^{N'-K'}  \left(\begin{array}{ccc} N' & 2 & N \\ -K' & q & K' \end{array}\right) \left(T_{-\alpha}\right)^2_q\bigg],
\end{eqnarray}
\end{widetext}
where $\Gamma$ denotes the symmetry of the nuclear spin wavefunction, $a_F$ is the Fermi contact parameter, and $T$ the dipole-dipole interaction tensor.

In CaOCH$_3$, it was found that the only significant matrix elements arise from the $(T_{\pm1})^2_{\pm2}$ term~\cite{namiki1998CaOCH3}. The hyperfine matrix elements can thus be simplified as:
\begin{widetext}
\begin{eqnarray}
& & \langle N',K', S, J', I, \Gamma', F, m_F| \hat{H}_{hf} | N, K, S, J, I,\Gamma F, m_F\rangle  \nonumber \\
&=& \bigg[a_F \delta_{m_F' m_F} \delta_{F'F} \delta_{N'N} \delta_{K'K} (-1)^{N+S'+J'} (-1)^{J'+I +F'+1} \sqrt{(2J'+1)(2J+1)} \nonumber \\
& &\times \sqrt{S(S+1)I(I+1)(2I+1)}\left\{\begin{array}{ccc} I & J' & F \\ J & I & 1 \end{array}\right\} \left\{\begin{array}{ccc} S & J' & N \\ J & S & 1 \end{array}\right\}\bigg] \nonumber \\
& &-\bigg[ \delta_{m_F' m_F} \delta_{F'F} (-1)^{J'+I+F'} \sqrt{30} \sqrt{S(S+1)(2S+1)(2J'+1)(2J+1)(2N'+1)(2N+1)} \left\{\begin{array}{ccc} I & J' & F \\ J & I & 1 \end{array}\right\}\notag \\
& &\times\left\{\begin{array}{ccc} N' & N & 2 \\ S & S & 1 \\ J' & J &1 \end{array}\right\}
\sum_{\alpha=\pm 1} \langle I,\Gamma' || I_\alpha ||I, \Gamma \rangle \sum_{q=\pm2} (-1)^{N'-K'}  \left(\begin{array}{ccc} N' & 2 & N \\ -K' & q & K' \end{array}\right) \left(T_{-\alpha}\right)^2_q\bigg],
\end{eqnarray}
\end{widetext}
where $(T_{\pm1})^2_\pm{2} = (T_{bb}-T_{cc})/\sqrt{24}$.

We note that for the remaining terms, only matrix elements of $\langle  E_{\pm} I ||T^1(I_{\mp})|| E_{\mp} I\rangle$ are non-zero. They are given by
\begin{equation}
\langle  E_{\pm} I ||T^1(I_{\mp})|| E_{\mp} I\rangle = -2\sqrt{I (I+1)(2I+1)}
\end{equation}

The hyperfine matrix elements are thus:
\begin{widetext}
\begin{eqnarray}
& &\langle N',K', S, J', I, \Gamma', F, m_F| \hat{H}_{hf} | N, K, S, J, I,\Gamma F, m_F\rangle \nonumber \\
&=&\bigg[a_F \delta_{m_F' m_F} \delta_{F'F} \delta_{N'N} \delta_{K'K} (-1)^{N+S'+J'} (-1)^{J'+I +F'+1} \sqrt{(2J'+1)(2J+1)} \times \nonumber \\
& &\times \sqrt{S(S+1)I(I+1)(2I+1)}\left\{\begin{array}{ccc} I & J' & F \\ J & I & 1 \end{array}\right\} \left\{\begin{array}{ccc} S & J' & N \\ J & S & 1 \end{array}\right\}\bigg] \nonumber \\
& & +2 \bigg[\delta_{m_F' m_F} \delta_{F'F} (-1)^{J'+I+F'} \sqrt{30} \sqrt{S(S+1)(2S+1)(2J'+1)(2J+1)(2N'+1)(2N+1)} \left\{\begin{array}{ccc} I & J' & F \\ J & I & 1 \end{array}\right\}\notag \\
& &\times \left\{\begin{array}{ccc} N' & N & 2 \\ S & S & 1 \\ J' & J &1 \end{array}\right\}\sqrt{I (I+1)(2I+1)} \sum_{\alpha=\pm 1} \delta_{\alpha,\Gamma} \delta_{\bar{\Gamma}', \Gamma} \sum_{q=\pm2} (-1)^{N'-K'}  \frac{1}{\sqrt{24}} \left(\begin{array}{ccc} N' & 2 & N \\ -K' & q & K' \end{array}\right) (T_{bb}-T_{cc})\bigg]
\end{eqnarray}
\end{widetext}
In practice, for $|K|=1$, one can drop the indices of $\Gamma$, since $|K=\pm\rangle$ states always occur with $|E_{\mp}\rangle$ spin states. The first term is analogous to a single nucleus of spin $I=1/2$. The second term breaks the degeneracy between $K=\pm 1$ states of identical $m_F$, and gives rise to resolvable $K$-doublets. The spacing of these $K$-doublets is thus set by the the size of $T_{bb}-T_{cc}$.

\section{Electric and Magnetic field dependence}
The electric and magnetic field dependence can be computed by diagonalizing the molecular Hamiltonian including terms correspond to external electric and magnetic fields. First, the Stark Hamiltonian $\hat{H}_S = -T^1(\mathbf{d}) \cdot T^1(\mathbf{E})$, where the matrix elements for $T^1_p(\mathbf{d})$ are given by

\begin{widetext}
\begin{eqnarray}
 & &\langle N',K',S',J',I',F',m_F'|T^1_p(\mathbf{d})  |N,K,S,J,I,F,m_F \rangle  \nonumber \\
&=& -d_0\delta_{I' I} \delta_{S' S} (-1)^{F'-m_F'} \threeJ{F'}{-m_F'}{1}{p}{F}{m_F} (-1)^{F+J'+1+I'}\sqrt{(2F+1)(2F'+1)}\sixJ{J}{F}{I'}{F'}{J'}{1} \nonumber \\
& & (-1)^{J+K'+1+S'} \sqrt{(2J+1)(2J'+1)} \sixJ{N}{J}{S'}{J'}{N'}{1} \sqrt{(2N'+1)(2N+1)}\threeJ{N'}{-K'}{1}{0}{N}{K} 
\end{eqnarray}
\end{widetext}

Next, we have the Zeeman Hamiltonian $\hat{H}_z= \hat{H}_{z,S} + \hat{H}_{z,I}$, with the two terms corresponding to the coupling of the electron spin and the nuclear spin to external magnetic field. The first term can be written in terms of spherical tensors as
\begin{eqnarray}
\hat{H}_{z,S} & =& g_S \mu_B T^1(\mathbf{S}) \cdot T^1(\mathbf{B})  \nonumber \\
& =& g_S \mu_B \sum_p (-1)^p T^1_p(\mathbf{S})T^1_{-p}(\mathbf{B}),
\end{eqnarray}
where the matrix elements for $T^1_p(\mathbf{S})$ are given by
\begin{widetext}
\begin{eqnarray}
& & \langle N',K',S',J',I',F',m_F'|T^1_p(\mathbf{S})  |N,K,S,J,I,F,m_F \rangle \nonumber \\
&=&\delta_{I'I} \delta_{N' N}\delta_{K',K} \delta_{S' S}(-1)^{F'-m_F'} \threeJ{F'}{-m_F'}{1}{p}{F}{m_F} (-1)^{F+J'+1+I'}\sqrt{(2F'+1)(2F+1)}\sixJ{J}{F}{I'}{F'}{J'}{1} \nonumber \\
& & (-1)^{J'+N'+1+S}\sqrt{(2J'+1)(2J+1)}\sixJ{S}{J}{N'}{J'}{S'}{1} \sqrt{S(S+1)(2S+1)} 
\end{eqnarray}
\end{widetext}
The nuclear spin term $\hat{H}_{z,I}$ is given by
\begin{eqnarray}
\hat{H}_{z,I} & =& -g_N \mu_N T^1(\mathbf{I}) \cdot T^1(\mathbf{B})\nonumber \\ 
& =& -g_N \mu_N \sum_p (-1)^p T^1_p(\mathbf{S})T^1_{-p}(\mathbf{B}),
\end{eqnarray}
where $T^1_p(\mathbf{I})$ is given by
\begin{widetext}
\begin{eqnarray}
 & &\langle N',K',S',J',I',F',m_F'|T^1_p(\mathbf{I})  |N,K,S,J,I,F,m_F \rangle  \nonumber \\
&=&\delta_{J'J} \delta_{I'I}\,(-1)^{F'-m_F'} \threeJ{F'}{-m_F'}{1}{p}{F}{m_F}(-1)^{F'+J'+1+I}\sqrt{(2F'+1)(2F+1)}\sixJ{I}{F}{J'}{F'}{I'}{1}\nonumber \\
& &\times\sqrt{I(I+1)(2I+1)} 
\end{eqnarray}
\end{widetext}

\section{Estimates of Experimental Parameters and Technical Imperfections}
\subsection{Stray Magnetic fields}
Due to the single unpaired electron in the $^2A_1$ ground state manifold, alkaline earth monomethoxides are sensitive to magnetic fields on the order of $\sim 1$\,MHz/G. But as discussed, the resolvable $K$-doublets naturally produce states that are magnetically insensitive at zero magnetic field. The leading order energy shift is quadratic in field. For magnetic field noise at the $10\,\mu\text{G}$ level (both orthogonal and parallel to quantization axis), which is achievable with magnetic shielding, the corresponding energy shifts are on the order of $h\times 0.01\,\text{Hz}$.

\subsection{Stray Electric Fields}
In our Monte-Carlo simulation, a DC electric field of 100\,V/cm is applied to align the molecules. AC noise in the applied electric field can be eliminated with filtering. The relative stability of this field can be controlled to the $\sim 1\times 10^{-7}$, which requires stabilization of voltages at the $10\,\mu\text{V}$ level for field plates spaced $\sim 1$ cm apart. The $\ket{e}$ state is most sensitive to instability in the electric field, since it has a large lab frame dipole moment. With $10^{-6}$ stability in the applied field, the corresponding shift in the energy of $\ket{e}$ is on the order of a few Hz.

We model locally fluctuating electric fields as a consequence of patch potentials induced by atoms or molecules adsorbed to nearby surfaces. These patch potentials produce two effects that are well-studied for ion traps. Firstly, these patch potentials can drive unintended transitions between rotational states via fluctuating AC electric fields. From studies of surfaces with ions, it has been found that the noise power of electric fields falls with distance from a surface $z$ as $z^4$ ~\cite{daniilidis2011fabrication,safavi2011microscopic}. Using a typical distance of $z\sim 1$\,cm, we expect the noise power $S(\omega)$ at frequency $\omega$ to be on the order of $\sim 10^{-12}\omega$  (V/m)$^2$. Using Fermi's Golden Rule, one obtains the rate of incoherent transitions due to patch potentials as:
\begin{equation}
	\Gamma_r=\frac{2\pi}{\hbar^2}d^2S(\omega_0)
\end{equation}
where $d$ is the dipole moment and $\omega_0$ is the frequency spacing between rotational levels. Taking $d$ to be $0.8$ Debye, we find the rate $\Gamma_r$ to be $4\times 10^{-3}\,\text{s}^{-1}$. Thus, the effect of patch potentials on driving incoherent transitions is negligible for gate times of $\sim 1\,\text{ms}$.

The second effect from patch potentials is decoherence due to fluctuating DC Stark shifts.  As the energy of a state with dipole moment $d$ couples to electric fields, energy shifts can lead to accumulated phase errors in a CNOT gate.  It is known that the noise spectrum of patch potentials becomes flat at frequencies below $\sim 1-10$\,MHz~\cite{safavi2011microscopic}. Using a realistic gate time of $\sim 1$\,ms, one only need to consider frequencies up to $\sim 1$\,kHz.  Using, a distance of $\sim 1$ cm from any surfaces, the integrated electric field noise from DC to $1$\,kHz is $10^{-8}$\,V/m. Thus, the resulting accumulated phase errors for a gate time of 1\,ms is on the order of $10^{-6}$. These estimates indicate that electric fields from patch potentials should not be a limiting factor. They are in fact negligible compared to the precision with which the applied DC electric field can be controlled.

\subsection{Variations in Rabi frequency and Optical Trap power}
Intensity variations of the microwave drives can lead to variations in the Rabi frequencies. In the simulation, the Rabi frequency is stabilized to $10^{-2}$, easily achievable in an experiment. Similarly, intensity fluctuations in the addressing lasers lead to fluctuations in the AC Stark shifts. In the far-detuned limit and in the limit of largely diagonal Frank-Condon factors, we can estimate the scalar shift $U$ with the expression:
\begin{equation}
   U \approx \frac{1}{\hbar}\bigg[\frac{\pi c^2}{\omega_{A}^3}\frac{\Gamma_{A}}{\Delta_A}+\frac{\pi c^2}{2\omega_{B}^3}\frac{\Gamma_{B}}{\Delta_B}\bigg]I
\end{equation}
where $I$ is the intensity, and the scalar shift is taken to be over the $\tilde{A}$ and $\tilde{B}$ states. The factors before the $\tilde{A}$ and $\tilde{B}$ terms arise from sum rules in the far-detuned limit. The excited state linewidth is taken to be $\Gamma_{A/B}\sim (35\,\text{ns})^{-1}$ based on spectroscopy of the $\tilde{A}\to\tilde{X}$ transition on CaOH.

We propose to use a blue-detuned trap. In this case, the variation in energy due to fluctuating intensities is proportional to leading order $k_B T$ plus the zero-point motional energy due to the trapping potential. Our calculations indicate that the $\ket{0}$ and $\ket{1}$ states are well-matched in AC stark shifts, while the $\ket{e}$ state differs by $30\%$ in the far-detuned limit with a single frequency optical trap. We thus estimate that the mean energy difference between $\ket{1}$ and $\ket{e}$ will shift by $0.3k_B T$, which is 6\,kHz at $T=1\,\mu\text{K}$. Stabilization at the $10^{-3}$ level, which is easily achievable, would make this contribution negligible to the dipolar energy shift. Stabilization to $10^{-4}$ would further limit fluctuations in the energy shift to the $1\,\text{s}^{-1}$ level. 
The use of magic trapping conditions will further suppress sensitivity to optical power fluctuations by many orders of magnitude. The issue of decoherence due to differential AC Stark shifts is not considered here, and will likely require the use of magic wavelength traps.

\subsection{Decoherence from Collisions and from Intrinsic Lifetime of Molecular states}
A separate source of decoherence are collisions. We propose to operate in the collisional blockade regime, where each lattice site is occupied by no more than 1 molecule. Except for long-range electric dipolar interactions, molecular interactions are negligible even when separated by a lattice spacing ($0.5\mu\text{m}$). The remaining collisions that can lead to decoherence is collisions with background gas. Since the background gas particles have average kinetic energy much larger than the trap depths used, molecules will be lost from the trap in a collision. The rate of background gas collision is determined by the vacuum pressure. Typical background gas-limited lifetimes in ultracold experiments are on the order of $100$\,s, and should not be a limitation in our proposed setup. The most likely limitation is the intrinsic lifetimes of monomethoxides, which is estimated to be on the order of seconds.

\subsection{Decoherence due to Spontaneous Emission}
Decoherence can also be caused by spontaneous emission induced by the optical lattice light. For far detuned optical traps, the rate of spontaneous emissions relative to trap depth is given by $\Gamma/(\hbar \Delta)$ where $\Gamma$ is the excited state linewidth and $\Delta$ is the detuning from resonance.  Using estimated values of $\Gamma = 2\pi\times 10$\,MHz for the linewidth, a trap depth of $U = k_B \times 0.5 \,\text{mK}$, and a detuning of $2\pi\times 100$\,THz, the expected photon scattering rate is $\sim 1\,\text{s}^{-1}$.  
Although the trap lifetime is longer by another factor of $\sim 10^3$, since each scattering event only increases the energy by $\sim 1$ $\mu$K, each spontaneous scattering event can change the internal state. If the final internal state is on of the qubit states, this gives rise to spin flip errors. If the final internal state is not one of the qubit states, this is equivalent to trap loss. The latter is the dominant effect, and is expected to occur at a similar rate to spontaneous scattering. With the stated parameters, we thus expect a decoherence rate of $\sim 1\,\text{s}^{-1}$.

\section{Monte-Carlo Simulation}
\subsection{Description of Master Equation Approach}
To model the evolution of the molecule, we follow a master equation approach. Each molecule is reduced to a three-level system, with the basis $\{\ket{e}, \ket{1}, \ket{0}\}$, where we integrate all coherent and incoherent effects through the master equation.  By extension, a tensored 9$\times$9 system describes the combined two-qubit gate.  In both cases, the master equation has the general form, where $\rho$ refers to the qubit density matrix, $H_{c/t}$ the rotating wave Hamiltonian, and $\mathcal{L}_{c/t}$ the Louvillian operator.
\begin{equation}
	\frac{\partial \rho_{c/t} }{\partial t}=-\frac{i}{\hbar}[H_{c/t}, \rho_{c/t}]+\mathcal{L}_{c/t}
\end{equation}
When solving for the dynamics of a single qubit, both $H_{c/t}$ and $\mathcal{L}_{c/t}$ are simply the corresponding 3$\times$3 operators.  The 3$\times$3 rotating wave Hamiltonian is defined as:
\begin{equation}
	H_{c/t}=\begin{pmatrix}
	\Delta_{e1} & \Omega_{e1} & 0\\
    \Omega_{e1} &  \Delta_{10} & \Omega_{10}\\
    0 & \Omega_{10} & 0
	\end{pmatrix}
\end{equation}
where $\Omega_{ab}$ refers to the Rabi frequency of the transition from $\ket{a}\to \ket{b}$ and $\Delta_{ab}$ is the quadrature sum of electric and magnetic field instabilities as well as Doppler shifts from thermal motion within the trap that lead to an effective detuning. 

Similarly, we define the 3$\times$3 Liouvillian operator as:
\begin{equation}
	L_{c/t}=\begin{pmatrix}
	0 & -\frac{\gamma_{e1}}{2}\rho_{e1} & -\frac{\gamma_{e0}}{2}\rho_{e0} \\
    -\frac{\gamma_{e1}}{2}\rho_{e1}  & 0 & -\frac{\gamma_{10}}{2}\rho_{10} \\
    -\frac{\gamma_{e0}}{2}\rho_{e0}  & -\frac{\gamma_{10}}{2}\rho_{10}  & 0
	\end{pmatrix}
\end{equation}
where $\rho_{ab}$ is the corresponding matrix term in the three-level density matrix for a single qubit.  The off-diagonal decoherence rate terms $\gamma_{ab}$ are calculated, similar to before, as the quadrature sum of E-field and B-field fluctuations; $\gamma_{ab}=\sqrt{\Delta_E^2+\Delta_B^2}$, where $\Delta_E$ corresponds to the patch potential fluctuation, and $\Delta_B$ is set at $1\times 10^{-6}$ Gauss.\par 
To describe a CNOT gate, the 3$\times$3 operators are replaced by their tensored 9$\times$9 counterparts, where we adopt the notation $H_{ct}$ and $\mathcal{L}_{ct}$ to denote the tensored operators. The full Hamiltonian $H_{ct}$ for two qubits is built from tensoring the individual rotating wave Hamiltonians for the target and control molecules, as well as adding in the dipole-dipole interaction:
\begin{equation}
	H_{ct}=H_c\otimes I_t + I_c \otimes H_t +\hbar U_B \begin{bmatrix}
	\mathbf{0}_{8} & 0\\
    0 & 1
	\end{bmatrix}\end{equation}
where $I_{c/t}$ is the identity operator acting on the control (c) and target (t) subspace respectively, $U_B$ is the dipolar blockade energy, and $\mathbf{0}_8$ is an 8$\times$8 zero matrix. Similarly, the Liouvillian operator, which incorporates the off-diagonal dephasing terms, is expressed as $\mathcal{L}_{ct}=\mathcal{L}_c\otimes I_t + I_c \otimes \mathcal{L}_t$, where  $\mathcal{L}_{c/t}$ refer to the 3$\times$3 Liouvillian operators. Simulations were performed using the master equation solver in the python package Qutip \cite{johansson2012qutip, johansson2013qutip}.

\subsection{Modeling Electric and Magnetic Field Variations}
\begin{figure}
	\includegraphics[width=\columnwidth]{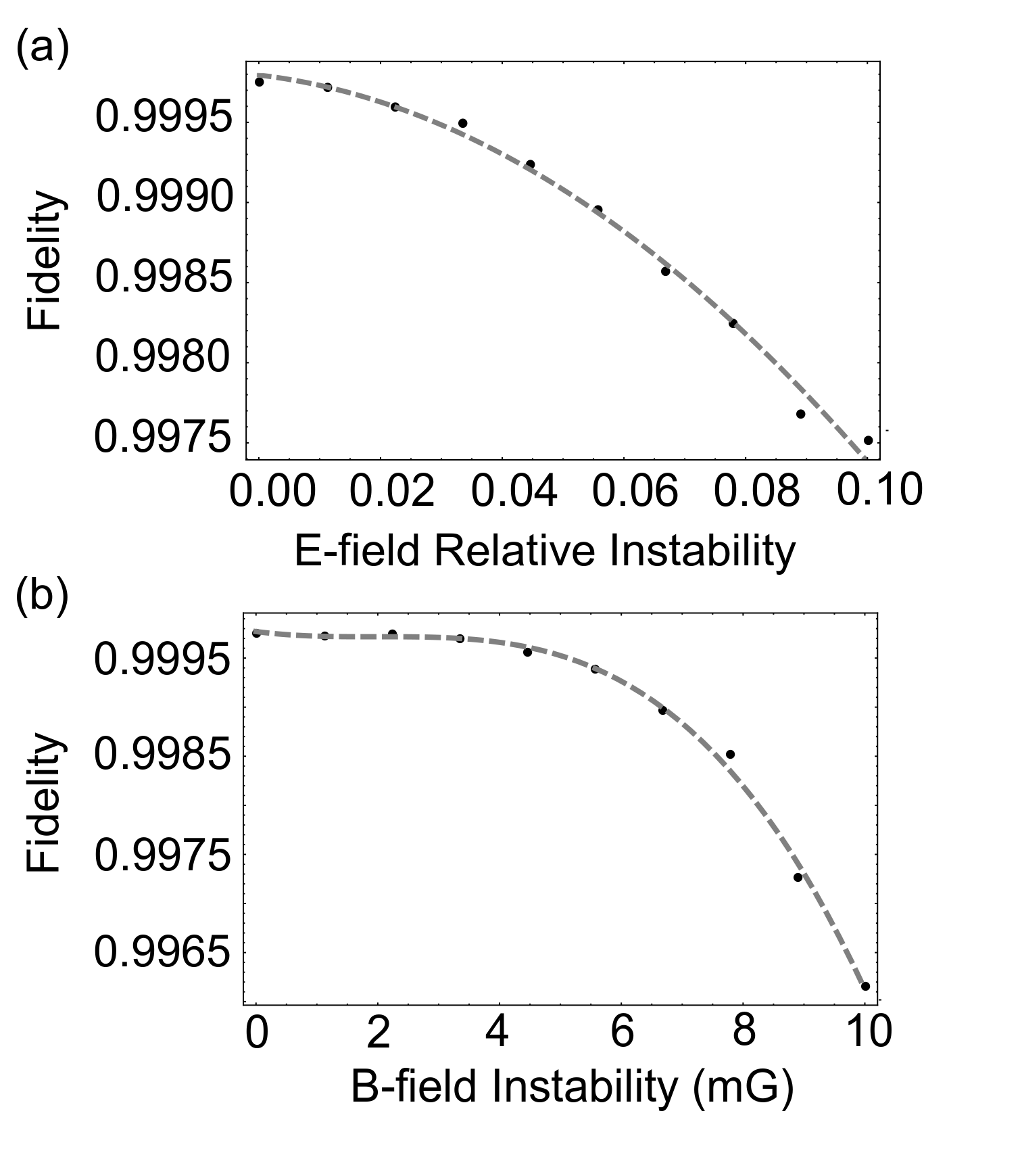}
	\caption{Fidelities of single qubit rotations $\ket{0}\to \ket{1}$ against (a) relative electric field instability and (b) magnetic field noise.}
	\label{fig:sf_scan_plotsone}
\end{figure}

\begin{figure}[h]
	\includegraphics[width=\columnwidth]{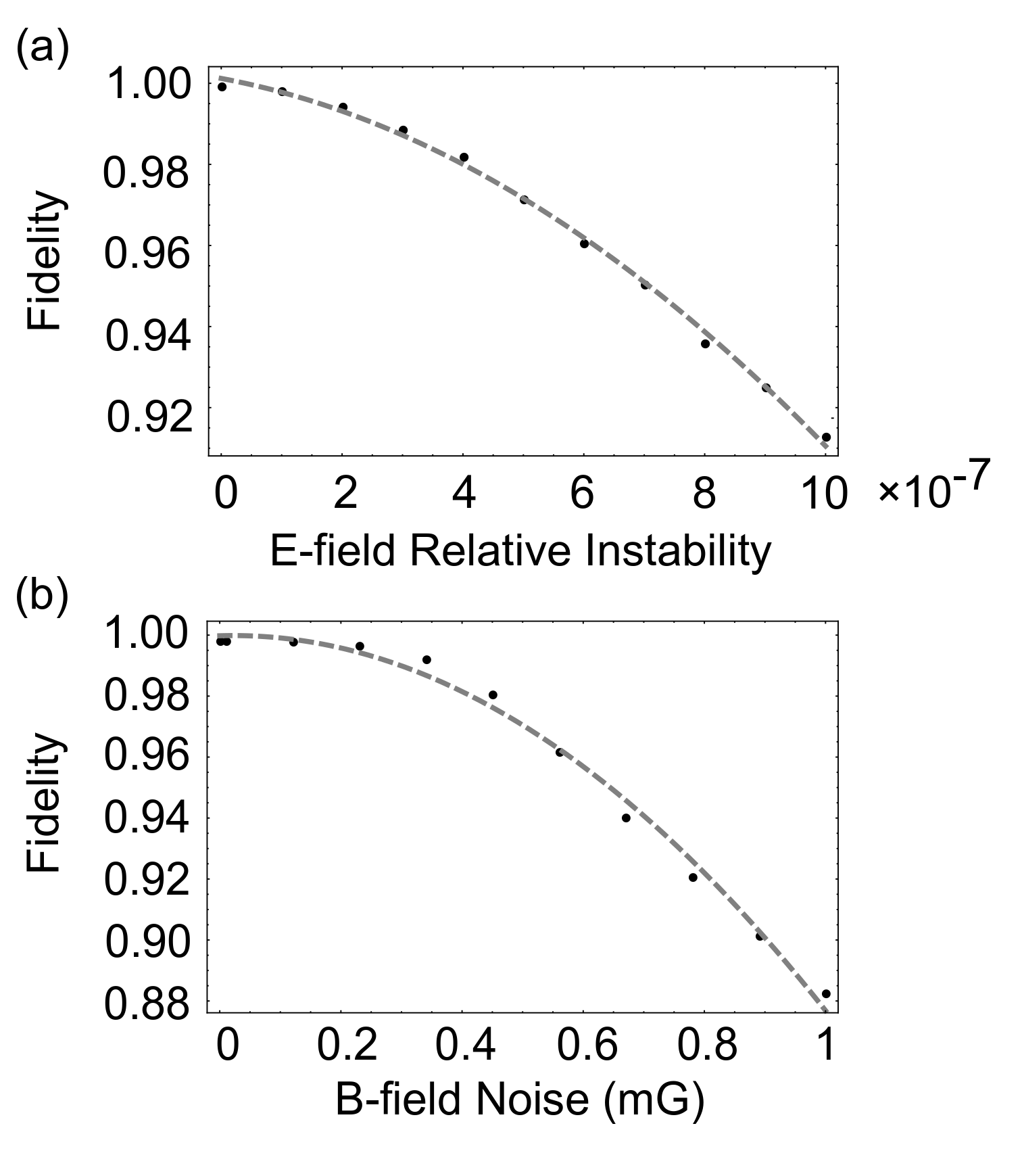}
	\caption{Fidelities of two qubit gate against (a) relative electric field instability and (b) magnetic field noise.}
	\label{fig:sf_scan_plotstwo}
\end{figure}

For modeling the effects of field instabilities on fidelities, we use the numerically calculated Stark and Zeeman shifts of CaOCH$_3$.  We simulate single (Fig. \ref{fig:sf_scan_plotsone}) and 2-qubit operations (Fig. \ref{fig:sf_scan_plotstwo}) with different field parameters, while holding all other error terms constant. We find that our current scheme is limited by electric and magnetic-field fluctuations at the $10^{-3}$ level under currently available technologies. Future improvements in control of environmental conditions could potentially allow lower errors to be achieved. 


\end{document}